\documentclass{eptcs}
\providecommand{\event}{GANDALF 2018} 

\usepackage{tikz}
\usepackage{amsmath,amssymb,stmaryrd,xspace,ifthen,wasysym,paralist}
\usepackage{algorithm, algorithmicx,calc,tkz-graph}
\usepackage{algpseudocode}
\usepackage[utf8x]{inputenc}
\usetikzlibrary{arrows,automata,shapes}

\usepackage{multicol}

\usepackage{pgfplotstable}
\usepackage{pgfplots}
\usepgfplotslibrary{units}
\def\nat{\mathbb{N}}

\usetikzlibrary{arrows,automata,shapes}

\newtheorem{defi}{Definition}
\newtheorem{theo}{Theorem}
\newtheorem{prop}{Proposition}
\newtheorem{lemm}{Lemma}
\newtheorem{coro}{Corollary}
\newtheorem{exam}{Example}

\newenvironment{definition}{\begin{defi} \it }{\end{defi}}
\newenvironment{example}{\begin{exam} \rm }{\end{exam}}

\newcommand{\qed}{\ensuremath{\hfill\Box}}
\newcommand{\pre}[2]{\ensuremath{\textit{pre}_{#1}(#2)}}
\newcommand{\post}[1]{\ensuremath{\textit{post}(#1)}}

\newcommand{\myomit}[1]{}

\newcommand{\Vars}{\ensuremath{\vec{x}}\xspace}

\newcommand{\pnot}[1]{\bar{#1}}

\newcommand{\player}{\ensuremath{\alpha}}

\newcommand{\attrsym}{\ensuremath{\textit{Attr}}}
\newcommand{\attr}[3][]{\ensuremath{\attrsym^{#1}_{#2}(#3)}}

\newcommand{\even}{0\xspace}
\newcommand{\odd}{1\xspace}

\newcommand{\ie}{\emph{i.e.}\xspace}
\newcommand{\eg}{\emph{e.g.}\xspace}
\newcommand{\viz}{\emph{viz.}\xspace}

\newcommand{\oftype}{{:}}

\begin{document}

\title{A Comparison of BDD-Based Parity Game Solvers}
\author{Lisette Sanchez, Wieger Wesselink and Tim A.C. Willemse
\institute{Eindhoven University of Technology, The Netherlands}
}
\date{}
\def\titlerunning{A Comparison of BDD-Based Parity Game Solvers}
\def\authorrunning{Lisette Sanchez, Wieger Wesselink and Tim A.C. Willemse}

\maketitle

\begin{abstract}
Parity games are two player games with omega-winning conditions,
played on finite graphs.  Such games play an important role in
verification, satisfiability and synthesis. It is therefore important
to identify algorithms that can efficiently deal with large games
that arise from such applications.  In this paper, we describe our
experiments with BDD-based implementations of four parity game
solving algorithms, \viz Zielonka's recursive algorithm, the more
recent Priority Promotion algorithm, the Fixpoint-Iteration algorithm
and the automata based APT algorithm. We compare their performance
on several types of random games and on a number of cases taken
from the Keiren benchmark set.

\end{abstract}

\section{Introduction}
\label{sec:introduction}

Parity games \cite{EJ:91,McN:93,Zie:98} are infinite duration games
played by two players on a finite directed graph. Each vertex in
the game graph is owned by one of the two players and vertices are
assigned a colour, or \emph{priority}.  The game is played by pushing
a token along the edges in the graph; the choice to which
vertex the token is to move next is decided by the player owning
the vertex currently holding the token. A parity condition
determines the winner of this infinite play. A vertex in the game
is won by the player that has a strategy for which, no matter how
the opponent plays, every play from that vertex is won by her; the
winner of each vertex is uniquely determined~\cite{McN:93}.

The parity game solving problem is to compute the set of vertices
won by each player. This problem is known to be in
$\text{UP}\cap\text{coUP}$, a class that indicates that the problem
is unlikely to be hard for NP.  While subexponential solving algorithms
have been devised, despite the (deceptive) simplicity
of the game and a continuous research effort, no polynomial time
algorithm for solving parity games has been found.  However, only
recently the problem was shown to be solvable in \emph{quasi-polynomial
time}~\cite{Calude,JurdzinskiL17,FearnleyJS0W17}.

From a practical viewpoint, parity games are interesting since they
underlie typical verification, satisfiability and synthesis problems,
see~\cite{CPPW:07,EJ:91,AVW:03} and the references therein. For
instance, in the mCRL2 toolset~\cite{CranenGKSVWW13}, parity games,
derived from parameterised Boolean equation systems~\cite{GrooteW05},
are used to solve modal $\mu$-calculus model checking problems for
reactive systems. Games originating from such model checking problems
often contain a large number of vertices (typically exceeding $10^6$),
whereas the number of distinct priorities in the games is typically
very small (often at most 3). Consequently, these applications
require algorithms and techniques that can efficiently deal with
such magnitudes.

Most parity game solving algorithms fall in one of two categories:
`strategy identification' (SI) algorithms and `dominion identification'
(DI) algorithms.  Algorithms from the SI category directly compute
the winning strategies for both players; \eg, by means of policy
iteration or by maintaining some statistics about the plays
that can emerge from a vertex.  The recent quasi-polynomial
algorithms~\cite{Calude,JurdzinskiL17,FearnleyJS0W17} all fall in
this category, but also classical algorithms such as Jurdzi\'nski's
\emph{small progress measures} algorithm~\cite{Jur:00}, and, to
some extent, the Fixpoint-Iteration (FI) algorithm~\cite{BruseFL14},
which is closely related to the small progress measures algorithm,
see \emph{ibid}.  The DI category of algorithms proceed by (recursively)
decomposing the game graph in \emph{dominions}: small subgraphs
that are won by a single player and from which the opponent cannot
escape. A classical exponent of the latter category is Zielonka's
\emph{recursive algorithm}~\cite{Zie:98} but also the recently
introduced Priority Promotion (PP) algorithm~\cite{BenerecettiDM16}.
While DI algorithms typically have a worst-case running time
complexity that is theoretically less attractive than that of SI
algorithms, in practice, the SI are significantly outperformed by
DI algorithms~\cite{Dijk}.

Since parity games that originate from practical verification
problems can become quite large, it is natural to study whether
existing algorithms can be implemented efficiently using symbolic
representations of the game graph, \ie, using Binary Decision Diagrams
(BDDs). While the size of BDDs may depend very much on the BDD
variable ordering, when the compression ratio (the number of BDD
nodes versus the number of satisfying assignments) is favourable,
BDDs can be used to concisely represent large graphs. DI algorithms
typically exploit set-based operations, which can be implemented
straightforwardly and efficiently using BDDs.  The same applies to
SI algorithms such as the FI algorithm~\cite{BruseFL14} and the
automata-based APT algorithm~\cite{StasioMPV16}.

In this paper, we describe and study four set-based algorithms,
\viz Zielonka's algorithm, the PP algorithm, the FI implementation
and the APT algorithm.  In particular, we reassess the conclusions
from~\cite{Dijk} in which Zielonka's algorithm and the PP algorithm
are shown to have similar performance when computing with 
explicit representations of the game graph. Moreover, we revisit the
observation of~\cite{StasioMPV16} that for games in which the number
of vertices is exponentially larger (given a sufficiently large
enough base) than the number of distinct priorities in the game,
the APT algorithm significantly outperforms Zielonka's algorithm.
Like the aforementioned works, we base our observations on random
games, which we convert to BDDs using a binary encoding. Such an
encoding typically yields BDDs with a poor compression ratio. While
this does not reflect scenarios in which BDDs excel, this does yield
important insights into which symbolic algorithms are more sensitive
to BDDs with poor compression ratios.  Finally, we compare the
performance of all four algorithms on several of the larger cases
taken from the Keiren benchmark set~\cite{Keiren} of parity games
and demonstrate that our symbolic algorithms can solve games that
cannot be solved using explicit techniques.

We find that the symbolic Zielonka and PP algorithms 
perform mostly similarly; the observations of~\cite{Dijk}
thus seem to largely carry over to the symbolic setting. However, in
our setting
we cannot faithfully reproduce the observations of~\cite{StasioMPV16}
regarding the superior performance of the APT algorithm for favourable
vertex/priority ratios. While APT's performance improves
with more favourable ratios,
it (and also the FI algorithm) suffers
more than Zielonka and PP when handling graphs represented by BDDs
with a poor compression ratio. Yet, when the BDDs
remain small, both APT and FI often outperform Zielonka and PP, as
witnessed by the cases taken from the Keiren benchmark~set.\medskip

\noindent\emph{Related Work.}
BDD-based implementations for parity game solving have been studied
in the past. In~\cite{BakeraEKR08}, an algorithm that is inspired
by Zielonka's algorithm is used to solve planning problems, whereas
in~\cite{KantP14}, a symbolic version of Zielonka's algorithm is
implemented in an effort to solve model checking problems represented
by parameterised Boolean equation systems~\cite{GrooteW05}. We note
that the focus of the latter work is to extract symbolic parity
games, represented by \emph{List Decision Diagrams} (LDDs) underlying parameterised Boolean equation
systems using a reachability analysis and solving the thus constructed
parity games. Since our primary aim is to compare the performance
of various symbolic parity game solving algorithms, we have followed
a slightly different and less general route in our experiments on
model checking, converting all finite data types occurring in such
equation systems to Booleans.  This allows us to avoid a costly
reachability analysis. Obviously, our insights into which algorithms
can be used best to solve a given parity game can be combined with
the technique of~\cite{KantP14} to obtain faster end-to-end
verification tooling.
Another line of research has concentrated on giving set-based versions of classical algorithms
such as the small progress measures algorithm, the bigstep algorithm
and the subexponential dominion decomposition algorithm, see \eg~\cite{BustanKV04,ChatterjeeDHL17}. Symbolic algorithms
studied for more restricted types of games 
include~\cite{AlfaroF07,BerwangerCWDH09}.\medskip

\noindent\emph{Outline.} In Section~\ref{sec:parity_games},
we introduce parity games and the relevant concepts. The four algorithms
that we compare are then introduced and discussed in Section~\ref{sec:solvers}
and we describe how these can be implemented using BDD techniques in
Section~\ref{sec:implementation}. In Section~\ref{sec:experiments}, we describe
our experimental evaluation of our implementations and we finish with
conclusions in Section~\ref{sec:conclusions}.

\section{Parity Games}
\label{sec:parity_games}

\begin{figure}[h]
\centering
\begin{tikzpicture}[>=stealth']
\tikzstyle{every node}=[draw, inner sep=1pt, outer sep=1pt,minimum size=12pt];
  \node[shape=rectangle,label=above:{$u_1$}] (v1)               {\scriptsize 5};
  \node[shape=diamond,label=above:{$u_2$},minimum size=16pt] (v2) [right of=v1,xshift=30pt] {\scriptsize 6};
  \node[shape=diamond,label=above:{$u_3$},minimum size=16pt] (v3) [right of=v2,xshift=30pt] {\scriptsize 4};
  \node[shape=diamond,label=below:{$u_5$},minimum size=16pt] (v4) [below of=v1] {\scriptsize 1};
  \node[shape=rectangle,label=below:{$u_6$}] (v5) [right of=v4,xshift=30pt] {\scriptsize 2};
  \node[shape=rectangle,label=below:{$u_7$}] (v6) [right of=v5,xshift=30pt] {\scriptsize 3};
  \node[shape=rectangle,label=above:{$u_4$}] (v8) [right of=v3,xshift=30pt] {\scriptsize 2};
  \node[shape=rectangle,label=below:{$u_8$}] (v7) [right of=v6,xshift=30pt] {\scriptsize 3};

\path[->]
  (v1) edge (v2) 
  (v2) edge (v4) edge (v6)
  (v3) edge (v2)
  (v4) edge (v1) edge (v5)
  (v5) edge (v2) 
  (v6) edge (v5)
  (v6) edge (v3)
  (v6) edge (v7)
  (v8) edge (v3) edge (v6)
  (v7) edge [bend left] (v8)
  (v8) edge [bend left] (v7)
;
\end{tikzpicture}
\caption{A game with 6 different priorities,  3 vertices owned by player
$\even$ and 5 owned by player $\odd$.}
\label{fig:running_example}
\end{figure}
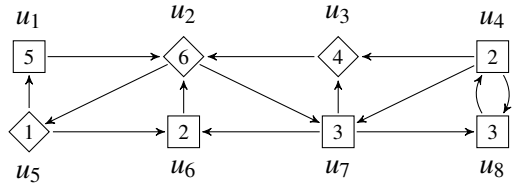

A parity game is an infinite duration game, played by players \emph{odd},
denoted by $\odd$ and \emph{even}, denoted by $\even$, on a directed,
finite graph. This game graph is formally defined as follows.
\begin{definition}
A parity game is a tuple $(V, E, p, (V_\even,V_\odd))$, where:
\begin{compactitem}
\item $V$ is a finite set of vertices, partitioned in a set $V_\even$ of
vertices owned by player $\even$, and a set of vertices $V_\odd$ 
owned by player $\odd$,
\item $E \subseteq V \times V$ is the edge relation; we assume that $E$ is
left-total, \ie, for all $v \in V$, there is some $w \in V$ such that
$(v,w) \in E$,
\item $p \oftype V \to \nat$ is a priority function that assigns
priorities to vertices.
\end{compactitem}
\end{definition}
We depict parity games as graphs in which diamond-shaped vertices represent
vertices owned by player $\even$ and box-shaped vertices represent vertices
owned by player $\odd$. Priorities, associated with vertices, are typically 
written inside vertices, see Figure~\ref{fig:running_example}.

\paragraph{Plays, Strategies and Winning.}

We write $v \to w$ whenever $(v,w) \in E$; $vE$ denotes the set $\{w \in V \mid v \to w\}$
and $Ew$ denotes the set $\{v \in V \mid v \to w \}$. 
Henceforth, $\player \in \{\even,\odd\}$ denotes an arbitrary player.
We write $\pnot{\player}$ for $\player$'s opponent; \ie, $\pnot{\even}=\odd$
and $\pnot{\odd}=\even$.  A sequence of vertices $v_1, \ldots, v_n$
is a \emph{path} if $v_m \to v_{m+1}$ for all $1 \leq m < n$.
Infinite paths are defined in a similar manner. The $n^\textrm{th}$ vertex
of a path $\pi$ is denoted $\pi_n$. 

A game starts by placing a token on vertex $v \in V$.  Players move
the token indefinitely according to a single simple rule: if the
token is on some vertex $v \in V_{\player}$, player $\player$ gets
to move the token to an adjacent vertex. We refer to the infinite
path through the graph, obtained this way, as a \emph{play}.
A play is won by player $\even$ iff the \emph{maximal} priority that occurs
infinitely often along that play is \emph{even}.

The moves of players $\even$ and $\odd$ are determined by their
respective \emph{strategies}.  A strategy for a player $\player$
determines, for a vertex $\pi_i \in V_{\player}$ the next vertex
$\pi_{i+1}$ that will be visited if the token visits $\pi_i$. In
general, a strategy is a partial function $\sigma \oftype V^* \times
V_{\player} \to V$ which, given a history of vertices visited by
the token and a vertex on which the token currently resides,
determines the next vertex. However, due to the positional determinacy
theorem for parity games~\cite{EJ:91} it suffices to consider
\emph{positional strategies}: partial functions of the form $\sigma
\oftype V_{\player} \to V$ that are compatible with $E$, \ie, for
all $v \in V_{\player}$ for which $\sigma$ is defined, it is the
case that $\sigma(v) \in vE$.   An infinite
path $\pi$ is compatible with a given strategy $\sigma$ if
for all vertices $\pi_i$ for which $\sigma$ is defined, we have
$\pi_{i+1} = \sigma(\pi_i)$.  A strategy $\sigma$ is \emph{closed} on a set of
vertices
$U$ iff every play compatible with $\sigma$ remains
within $U$. We say that a set of vertices $U$ is \emph{$\player$-closed}
iff $\player$ has a strategy that is closed on $U$.

A strategy $\sigma$ for player $\player$ is \emph{winning} from a
vertex $v$ if and only if $\player$ is the winner of every play
starting in $v$ that is compatible with $\sigma$.  A set of vertices
$W_{\player}$ is won by $\player$ if for each vertex $v \in
W_{\player}$, player $\player$ has some winning strategy from $v$.
Another consequence of the aforementioned positional determinacy theorem is that
the set of vertices $W_\even$ won by player $\even$ and the set of
vertices $W_\odd$, won by player $\odd$, forms a partition of the set $V$:
every vertex is won by exactly one player.
\begin{example}
Consider the parity game depicted in Fig.~\ref{fig:running_example}.
The vertices $u_1,u_2,u_3,u_5,u_6$ are won by player~$\even$ whereas 
vertices $u_4,u_7$ and $u_8$ are won by player $\odd$.  To see why vertices $u_4$ and $u_8$
are won by player~$\odd$, consider her strategy~$\sigma$ (that is \emph{closed} on $\{u_4,u_8\}$) defined by 
$\sigma(u_4) = u_8$ and $\sigma(u_8) = u_4$. Any play that is compatible with $\sigma$ infinitely often 
visits a dominating odd priority and is thus won by player~$\odd$.\qed

\end{example}

\paragraph{Subgames, Attractors and Dominions.}

For a game $G =( V, E, p, (V_\even,V_\odd) )$ and a set
$U \subseteq V$, we define the \emph{subgame} of $G$, denoted $G
\cap U$, as the maximal substructure $( V',E',p',(V'_\even,V'_\odd)
)$ that is obtained by restricting the graph $(V,E)$ to $U$
and imposing the restriction on the other elements of $G$, \ie, $V'
= U$, $E' = E \cap (U \times U)$, $p'(v) = p(v)$ for all $v \in U$,
and $V'_\even = V_\even \cap U$ and $V'_\odd = V_\odd \cap U$. If
the edge relation $E'$ of $G \cap U$ is again left-total, then the subgame
is again a parity game.  We furthermore use the abbreviation $G
\setminus A$ to denote the subgame $G \cap (V \setminus A)$.

Parity game solving can essentially be reduced to identifying or
computing appropriate subgraphs that are entirely won by a single
player and from which the opponent cannot escape. A subgame that
has this property for a player $\player$ is called an
$\player$-\emph{dominion}.  Technically, an $\player$-dominion is
a non-empty set of vertices $D_\player \subseteq V$ such that player
$\player$ has  a winning strategy, closed on $D_\player$, from \emph{all} vertices in the
set $D_\player$. 

The \emph{pre-image} $\pre{\player}{U}$ (or single-step
\emph{force}~\cite{StasioMPV16}) for a given non-empty set $U
\subseteq V$  and player $\player$ is defined as $ \pre{\player}{U}
= \{ v \in V_\player \mid vE \cap U \neq \emptyset \} \cup \{ v \in
V_{\pnot{\player}} \mid vE \subseteq U \} $.  The $\player$-attractor
into $U$, denoted $\attr{\player}{U}$, is the set of vertices for
which player $\player$ can force play into $U$.  Formally,
$\attr{\player}{U}$ is the least set $A \supseteq U$ to
which $\player$ can force play: $\attr{\player}{U} = \mu A. (U \cup
\pre{\player}{A} )$.
The \emph{confined} $\player$-attractor, denoted $\attr{\player}{T,U}$,
is defined analogously by limiting the set of vertices to those
from $T$. 
That is, it represents the subset
of vertices $T' \subseteq T$ from which $\player$ can force play
to $U$ while remaining in $T$.  We note that for arbitrary set $U
\subseteq V$, the subgame $G \setminus \attr{\player}{U}$ is again
a parity game; that is, the edge relation of the subgame is again
total. Furthermore, the $\player$-attractor into an $\player$-dominion
$D_\player$ is again an $\player$-dominion. Finally, observe that
the set $V \setminus \attr{\player}{U}$ is $\pnot{\player}$-closed.
\begin{example}
Reconsider the parity game from Fig.~\ref{fig:running_example}.
We already identified a closed strategy that is winning for player $\odd$
on the set of vertices $\{u_4, u_8\}$. So this set is a $\odd$-dominion. It is
not a maximal $\odd$-dominion: we can add $u_7$ to extend it. 
The $\even$-attractor into the set $\{u_2\}$ contains the
vertices $u_1$ (which is owned by $\odd$ but trivially attracted to $u_2$ since
it has but one successor vertex), $u_3,u_5$ and $u_6$. Vertex $u_4$ does not belong
to the $\even$-attractor into $\{u_2\}$ as player $\odd$ can choose to move to
$u_7$. For similar reasons, vertices $u_7$ and $u_8$ do not belong to the
$\even$-attractor into $\{u_2\}$. \qed
\end{example}

\section{Parity Game Solving Algorithms}
\label{sec:solvers}

We consider four set-based algorithms for solving parity games.
These algorithms will be introduced in some detail in the next subsections.

\subsection{Zielonka's Recursive Algorithm}\label{sec:zielonka}

The algorithm by Zielonka is a divide and conquer
algorithm that searches for dominions in subgames.  This algorithm
is essentially distilled from a constructive proof of the positional
determinacy of parity games by, among others, Zielonka~\cite{Zie:98}.
Despite the fact that the algorithm has a relatively bad theoretical
worst-case complexity (it runs in $\mathcal{O}(m n^d)$ where $n$
is the number of vertices, $m$ the number of edges and $d$ is the
number of different priorities in the game) and exponential worst-case
examples are known for various classes of special games~\cite{GazdaW13},
the algorithm remains among the most successful solvers for parity
games in practice, see~\cite{FriedmannL09} and the recent
exploration~\cite{Dijk}.

Zielonka's algorithm (see Algorithm~\ref{alg:zielonka}) constructs
winning regions for both players out of the solution of subgames
with fewer different priorities and vertices.  For this, the
algorithm relies on the fact that higher priorities in the game
dominate lower priorities, and that any forced revisit of these
higher priorities is beneficial to the player associated with the
parity of the priority.  For non-trivial games, the algorithm relies
on attractor set computations (see lines~8 and~10) to identify those
vertices that can be forced to visit the maximal priority;
the remaining vertices are then solved recursively (see line~9).

Assuming that the maximal priority is of parity $\player$, the
outcome of the first recursion yields a dominion for $\pnot{\player}$
in the subgame.  In case this dominion is $\pnot{\player}$-maximal
in the entire game, $\player$ wins all vertices outside the dominion
of player $\pnot{\player}$ (lines~11-12).  This can be seen as
follows: since the game restricted to $V\setminus B$ (where $B$ is computed
in line~10) is $\pnot{\player}$-closed, the opponent can choose to
stay within the subgame $W_\player$; choosing to do so means she will lose. So
the only option she has is to escape.  But the only escape she has
leads, via $A$, to the maximal priority which has the parity of $\player$.
Note that if $\player$ is forced to leave this maximal priority she
will again end up in the same subgame; any play that does so \emph{ad
infinitum} visits the maximal priority infinitely often and is
therefore won by $\player$.  In case $\pnot{\player}$'s dominion
is \emph{not} maximal, we can remove its $\pnot{\player}$-attractor
$B$ from the game and recursively solve the remaining subgame and
use its solution to construct the solution of the entire game (lines~14 and~15).

\begin{algorithm}[ht]
\caption{Zielonka's Algorithm}
\label{alg:zielonka}
\footnotesize
\begin{multicols}{2}
\begin{algorithmic}[1]
\Function{Zielonka}{$G$}
\If{$V = \emptyset$}
     \State $(W_\even,W_\odd) \gets (\emptyset,\emptyset)$
\Else
     \State $m \gets \max\{ p(v) ~|~ v \in V\}$
     \State $\player \gets m \mod 2$
     \State $U \gets \{v \in V ~|~ p(v) = m\}$
     \State $A \gets \attr{\player}{U}$
     \State $(W'_\even, W'_\odd) \gets \Call{Zielonka}{G\setminus A}$
     \State $B \gets \attr{\pnot{\player}}{W'_{\pnot{\player}}}$
     \If{$B = W'_{\pnot{\player}}$}
           \State $(W_{\player}, W_{\pnot{\player}}) \gets (A \cup W'_{\player}, B)$
     \Else
           \State $(W'_\even, W'_\odd) \gets \Call{Zielonka}{G\setminus B}$
           \State $(W_{\player},W_{\pnot{\player}}) \gets (W'_{\player},W'_{\pnot{\player}} \cup B)$
     \EndIf
\EndIf
\State \Return $(W_\even, W_\odd)$
\EndFunction
\end{algorithmic}
\end{multicols}
\end{algorithm}
Note that rather than
solving the entire game at once, one can first decompose a game into its strongly connected
components (SCCs) and first solve the bottom SCCs. The SCC decomposition can be integrated
tightly in the algorithm so that in each recursive call first an SCC decomposition is
performed. While this may sound expensive, in~\cite{GazdaW13} it is shown that this actually
allows the algorithm to run in polynomial time on many practically relevant classes of
special games such as solitaire and dull games for which the original algorithm might
require exponential time otherwise. 

\subsection{Priority Promotion}\label{sec:PP}

\newcommand{\esc}[3]{\ensuremath{\textit{esc}^{#1}_{#2}(#3)}}
\newcommand{\bep}[3]{\ensuremath{\textit{bep}^{#1}_{#2}(#3)}}

The recent \emph{Priority Promotion} (PP) algorithm~\cite{BenerecettiDM16}
starts, like Zielonka's recursive algorithm, with a game decomposition
that aims at identifying dominions for a given player. In contrast
to Zielonka's algorithm, however, the PP algorithm does not explicitly
solve subgames. Instead, within a fixed game, it uses a dominion
searcher that maintains a set of vertices, along with an updated
(promoted) priority mapping and zooms in on a dominion within that
set of vertices.  Once an $\player$-dominion is returned by the
dominion searcher,
the search for another dominion continues until the entire game is solved,
see Algorithm~\ref{alg:PP}.

\begin{algorithm}[ht]
\caption{Priority Promotion Solver}
\label{alg:PP}
\footnotesize
\begin{multicols}{2}
\begin{algorithmic}[1]
\Function{SolvePP}{$G$}
\State $(W_\even, W_\odd) \gets (\emptyset, \emptyset)$
\State $U \gets V$
\While{$U \neq \emptyset$}
\State $m \gets \max\{ p(v) \mid v \in U \}$
\State $(W'_\even,W'_\odd) \gets \Call{SearchDominion}{G \cap U,U,p,m}$
\State $(W_\even,W_\odd) \gets (W_\even \cup W'_\even, W_\odd \cup W'_\odd)$
\State $U \gets U \setminus (W'_\even \cup W'_\odd)$
\EndWhile
\State \Return $(W_\even, W_\odd)$
\EndFunction
\end{algorithmic}
\end{multicols}
\end{algorithm}

The main complexity and novelty of the PP algorithm lies in the way
it identifies a dominion. To this end, it relaxes the notion of a
dominion to a \emph{quasi} $\player$-dominion.  A quasi $\player$-dominion
is a set $U$ of vertices for which $\player$ has a strategy that
guarantees that every play that remains within $U$ is won by
$\player$, or that exits $U$ via the \emph{$\pnot{\player}$-escape}
of $U$. The $\player$-escape of a set $U$, denoted $\esc{\player}{G}{U}$
is the set of vertices from which $\player$ can force play out of
$U$ in a single move; \ie, 
$\esc{\player}{G}{U} = \pre{\player}{V \setminus U} \cap U$.
A quasi $\player$-dominion
$U$ in a game $G$ is said to be \emph{$\player$-closed} iff
$\esc{\pnot{\player}}{G}{U} = \emptyset$; otherwise it is said to
be $\player$-open. Note that an $\player$-closed quasi $\player$-dominion is
an $\player$-dominion. When the $\pnot{\player}$-escape set of a quasi 
$\player$-dominion only contains vertices with the highest priority in the
game, the $\player$ quasi-dominion is called an \emph{$\player$-region}.

By searching for quasi-dominions, the algorithm avoids the `hard'
problem of identifying the (set of) priorities that lead to a player
winning its dominion. Under certain conditions, these quasi-dominions
offer just enough guarantees that when composed, the result is a
dominion. Regions meet these conditions. The dominion searcher, see
Algorithm~\ref{alg:PP_search}, essentially traverses a partial order
of \emph{states} which contain information about the open regions
the searcher has computed up to that point. Once the searcher finds
a closed region it terminates and returns this dominion to the
solver. Conceptually the searcher assigns to each $\player$-region
it finds a priority that under-approximates the best value the
opponent $\pnot{\player}$ must visit when escaping from the
$\player$-region. A higher $\player$-region $U_1$ can then be merged
with a lower region $U_2$, yielding a new $\player$-region and
improving the under-approximation of $U_2$ by promoting its best
escape value.
\begin{algorithm}[h!]
\caption{Priority Promotion Dominion Searcher}
\label{alg:PP_search}
\footnotesize
\begin{multicols}{2}
\begin{algorithmic}[1]
\Function{SearchDominion}{$G,V_g,p_g,m_g$}
     \State $\player \gets m_g \mod 2$
     \State $U \gets \{v \in V_g \mid p_g(v) = m_g \}$
     \State $A \gets \attr{\player}{V_g,U}$
     \State $X \gets \esc{\pnot{\player}}{G}{A}$
     \If{$X =\emptyset$}
           \State $(W'_\player,W'_{\pnot{\player}}) \gets (\attr{\player}{A}, \emptyset)$
           \State \Return $(W'_\even,W'_\odd)$
     \Else
           \State $X \gets \esc{\pnot{\player}}{G \cap V_g}{A}$
           \If{$X \neq \emptyset$}
                     \State $p_g^* \gets p_g[A \mapsto m_g]$
                     \State $m_g^* \gets \max \{ p_g^*(v) \mid v \in V_g \wedge p_g^*(v) < m_g\}$
                     \State $V_g^* \gets V_g \setminus A$
           \Else
                     \State $m_g^* \gets \bep{\pnot{\player}}{p_g}{A}$
                     \State $p_g^* \gets (p \uplus p_g^{\ge m_g^*})[A \mapsto m_g^*]$
                     \State $V_g^* \gets \{v \in V \mid p_g^*(v) \le m_g^*\}$

           \EndIf
           \State \Return $\Call{SearchDominion}{G,V_g^*,p_g^*,m_g^*}$
     \EndIf
\EndFunction
\end{algorithmic}
\end{multicols}
\end{algorithm}

The searcher first checks whether the set of vertices $U$ that
dominate the current state induces a closed region $A$ within the
entire game; if so, the region is returned and the search is finished
(lines~2-8). If the region is open, the opponent $\pnot{\player}$
may escape.  When the region is open in the current state, she may
try to escape to some inferior priority within the subgraph; if the
region is closed in the current state, she can only escape to some
region in the larger game and end up in a priority that dominates
$m_g$.  

In case the region is open in the current state  (line~11-14), the
priority function of the current state is updated to set all vertices
in $A$ to the currently dominating priority $m_g$. This is achieved
by the update $p_g^* \gets p_g[U \mapsto m_g]$ which is defined as
$p_g^*(v) = m_g$ in case $v \in U$ and $p_g^*(v) = p_g(v)$ otherwise.
The new priority for the subgraph that will be explored is set to
the next largest priority in the graph (line~13) and $A$ is removed
from the subgraph (line~14). This new state is then explored
recursively (see line~20). In case the region is closed in the
current state (line~16-18), the opponent must escape (if she wants
to) to vertices in the larger game. The \emph{best escape priority}
she can force is the minimal priority among the set of vertices
that she can force to reach in a single step. This is given by the
function $\bep{\player}{p_g}{A}$ which yields the minimum priority
(according to $p_g$) of the set $\{w \in V\setminus A \mid \exists
v \in A \cap V_\player: w \in vE \}$.  The dominating priority $m_g$
is updated to $m_g^*$ to reflect this best escape priority (see
line~16) after which $p_g$ is updated to $p_g^*$. In this update,
all vertices in $A$ are set to priority $m_g^*$ while all vertices
with priorities exceeding $m_g^*$ remain unchanged.  All vertices
with priorities dominated by $m_g^*$ are reset to their original
value. This is achieved by the update in line~17. Here, $p_g^{\ge
m}$ yields the partial function that coincides with $p_g$ on the
(maximal) domain $V' \subseteq V$ for which $p_g(v) \ge m$ for all
$v \in V'$, and is undefined elsewhere. The \emph{update} $p \uplus
p_g$, for a partial function $p_g$ is then defined as $(p \uplus
p_g)(v) = p(v)$ in case $v \notin \textsf{dom}(p_g)$ and $p_g(v)$
otherwise. The subgraph of $G$ that is explored next is set to all vertices
with (promoted) priority no larger than $m_g^*$. This newly constructed
state is then again recursively explored (see line~20).

The PP algorithm resets all information in lower regions when
promoting vertices.  As observed in~\cite{BenerecettiDM16a,BenerecettiDM16b}, this can be improved
by only resetting regions of the opponent, or even a subset of
that. While this affects the performance for some
artificial worst-case examples, these improvements do not lead to
improvements on games stemming from practical applications, nor on
random games~\cite{Dijk}.

\subsection{The Fixpoint-Iteration Algorithm}\label{sec:FI}

The third algorithm we consider is the so-called \emph{Fixpoint-Iteration} (FI)
algorithm. It is essentially based on the reduction of the
parity game solving problem to the modal $\mu$-calculus model
checking problem. This transformation views a parity
game as a graph with state labels indicating the owner (by associating
a proposition \textsf{Even} or \textsf{Odd} to a state) and priority
of a vertex (by associating a proposition $\textsf{prio}_i$ to a
state representing a vertex with priority $i$), while the winning
regions $W_\even$ and $W_\odd$ of the game are characterised by a
so-called Walukiewicz $\mu$-calculus formula.

Rather than using a general-purpose modal $\mu$-calculus model checker 
for model checking a parity game, we here recall the algorithm given by
Bruse \emph{et al.}~\cite{BruseFL14}, see Algorithm~\ref{alg:FI}. 
This algorithm works directly on
a (pre-processed) parity game, essentially checking the following formula\footnote{Here, $\mu$ and $\nu$ are
the least and greatest fixpoint, respectively, $\diamond$ is the \emph{may} modality, $\Box$ is the
\emph{must} modality, $\textsf{Even}$, $\textsf{Odd}$ and $\textsf{prio}_i$ are propositions
and $\overline{\textsf{prio}}_i$ is the negation of $\textsf{prio}_i$.}
for games with least priority $0$, and a similar formula for games with
least priority $1$:
\[
\sigma X_{d-1} \dots \mu X_1. \nu X_0. ( ( \textsf{Even} \wedge \bigvee\limits_{i = 0}^{d-1} \diamond(\textsf{prio}_i \wedge X_i) )
                                    \vee ( \textsf{Odd} \wedge \bigwedge\limits_{i = 0}^{d-1} \Box(\overline{\textsf{prio}}_i \vee X_i) )
\]
The pre-processing guarantees that the parity game is \emph{compressed}:
\ie, it ensures that the set of priorities occurring in the game
constitute an integer interval with least priority $0$ or $1$.  Note
that converting a parity game to a compressed parity game can be
done without changing the solution to the game.
\begin{algorithm}[h!]
\caption{Fixpoint-Iteration algorithm for compressed parity games with least priority $0$ and
maximal priority $d-1$.
The algorithm for compressed parity games with least priority $1$ is analogous.}
\label{alg:FI}
\footnotesize
\begin{multicols}{2}
\begin{algorithmic}[1]
\Function{FI}{$G$}
      \Function{Diamond\_Box}{}
           \State $U_\even \gets\{v \in V_\even \mid \exists w \in vE: w \in X_{p(w)} \}$
           \State $U_\odd \gets \{v \in V_\odd \mid \forall w \in vE:~ w \in X_{p(w)} \}$
           \State \Return $U_\even \cup U_\odd$
     \EndFunction
     \For{$i \gets d-1, \dots, 0$}
           \State $X_i \gets \textbf{if} \text{ $i$ is even} \textbf{ then } V \textbf{ else } \emptyset$
     \EndFor
     \Repeat
           \State $X'_0, X_0 \gets X_0, \Call{Diamond\_Box}{\,}$
           \State $i \gets 0$
           \While{$X_i = X_i'$ and $i < d-1$}
                  \State $i \gets i+1$
                  \State $X_i', X_i \gets X_i, X_{i-1}$
                  \State $X_{i-1} \gets \textbf{if} \text{ $i$ is odd} \textbf{ then } V \textbf{ else } \emptyset$
           \EndWhile
     \Until{$i = d-1$ and $X_{d-1} = X'_{d-1}$}
     \State \Return $(X_{d-1}, V \setminus X_{d-1})$
\EndFunction
\end{algorithmic}
\end{multicols}
\end{algorithm}
The main loop of the algorithm evaluates the fixpoint-free part of
the Walukiewicz formula using the current values for the variables
$X_0\dots X_{d-1}$. The evaluation gives rise to new values for these
variables until a fixpoint is reached for all variables; at that
moment, variable $X_{d-1}$ contains all vertices that satisfy the
formula, and, by construction, are won by player $\even$.

Note that the re-initialisation of all variables $X_0\dots X_{d-1}$
can be optimised by exploiting monotonicity of the formula, reducing
the overall complexity from exponential in $d$ to exponential in
$\lceil d/2 \rceil$, see \eg~\cite{Seidl96}. This can be done by
first computing the smallest index $i$ for which $X_i$ is not yet
stable, and only resetting $X_{i-1},X_{i-3},\dots$ afterwards, thus
yielding an algorithm that is closely related to the small progress
measures algorithm, see~\cite{BruseFL14}. We have included this
optimisation in our implementation.

\subsection{The APT Algorithm}
\label{sec:APT}

The APT\footnote{The description of the algorithm in~\cite{StasioMPV16} seems
to contain several small mistakes and implementing it that way yields incorrect
solutions; our exposition is based on their actual
OCaml implementation in PGSolver.} algorithm, introduced by Kupferman
and Vardi~\cite{KupfermanV98}, is studied and implemented in~\cite{StasioMPV16}. This
algorithm solves parity games by solving the emptiness problem of
a corresponding alternating parity word automaton, which, in turn,
is solved by checking the emptiness of an equivalent weak alternating
word automaton.  Rather than explicitly constructing this weak
alternating word automaton, the APT algorithm, see Algorithm~\ref{alg:APT},
constructs this automaton implicitly, checking emptiness on-the-fly.

APT uses two disjoint sets of vertices $V_V$ (\emph{Visiting}
vertices) and $V_A$ (\emph{Avoiding} vertices) for solving a parity
game.  It declares a vertex \emph{Visiting} for a player just
whenever by reaching that vertex she can force a winning play; dually, 
a vertex is \emph{Avoiding} for a
player whenever by reaching that vertex she cannot force any winning
play. These sets derive their meaning from
a generalisation of the parity game winning condition, in which player
$\even$ wins a play as soon as $V_V$ is visited, player $\odd$ wins as
soon as $V_A$ is visited, and the traditional parity game condition applies
if neither $V_V$ nor $V_A$ are ever visited.  Furthermore, the algorithm keeps track of an
index $i$, representing the most significant priority in the game
for which the winners of the vertices with that priority have not yet
been decided for the given sets $V_V$ and $V_A$. 
\begin{algorithm}[h!]
\caption{The APT algorithm for compressed parity games with minimal priority $m$ and
maximal priority $d$.}
\label{alg:APT}
\footnotesize
\begin{multicols}{2}
\begin{algorithmic}[1]
\Function{APT}{$G$}
     \Function{Win}{$\player, i, V_V, V_A$}
           \State \Return $V \setminus \Call{FP}{\pnot{\player}, i, V_V, V_A} \textbf{ if } i \ge m \textbf{ else } \pre{\player}{V_V}$
     \EndFunction

     \Function{FP}{$\player,i,V_V,V_A$}
            \State $X,F \gets \emptyset, \{ v \in V \mid p(v) = i \}$
            \State $Y \gets \Call{Win}{\player,i-1,V_A \cup (F \setminus X),V_V \cup (F \cap X)}$
            \While{$Y \neq X$}
                  \State $X \gets Y$
                  \State $Y \gets \Call{Win}{\player,i-1,V_A \cup (F \cap X),V_V \cup (F \setminus X)}$
            \EndWhile
              
     \EndFunction
     \State $\player \gets (d+1) \mod 2$
     \State $X \gets \Call{Win}{\player, d, \emptyset,\emptyset}$
     \State \Return $(X,V \setminus X) \textbf{ if } \player = 0 \textbf{ else } (V \setminus X, X)$
\EndFunction
\end{algorithmic}
\end{multicols}
\end{algorithm}

\section{Implementing Parity Game Solvers Using BDDs}
\label{sec:implementation}

Using an efficient data structure to explicitly represent a graph,
one may be able to represent parity games up to $2^{25}$ vertices
on a 16Gb main memory machine.  Parity games encoding verification
problems of (software or hardware) systems easily require more
memory.  A symbolic representation of the game graph may then help
to sidestep this problem. Binary Decision Diagrams (BDDs) provide
just such a representation.
We assume that the reader is familiar with BDDs and the algorithms
manipulating these; for a comprehensive treatment we refer
to~\cite{Wegener00,DrechslerB98}. 

BDDs can be used to canonically (and often concisely) represent propositional
formulae, which in turn can be used to characterise sets. In our
setting we have to represent the following sets: the set of vertices
$V$ of a game, the edges $E$, the partition $(V_\even, V_\odd)$ of
$V$ and the priority function $p$. For most parity games stemming
from practical verification problems, the number of distinct
priorities in the game is small.  For that reason, it suffices to
represent $p$ as a mapping from priorities to sets of vertices.  For the computations
involved in the parity game solving algorithms of the previous
section we provide the key ingredients below; each operation is
described using an expression for the BDD-based implementation.

\newcommand{\BDD}[3][]{\ensuremath{\mathcal{#2}_{#1}{(#3)}}}

Henceforward we assume that $\Vars$ is the vector of Boolean variables used to
span the set of vertices $V$ of a given parity game; we assume $\iota$ is a total injective
mapping from $V$ to truth-assignments for $\Vars$. The $i$-th Boolean
variable of $\Vars$ is denoted $\Vars_i$; this notation extends to
other vectors of Boolean variables.  We represent a set $U \subseteq V$
by a propositional formula encoded as a BDD $\BDD{U}{\Vars}$ that
ranges over $\Vars$; as is standard, for a truth-assignment $\vec{u}$
for $\Vars$ we have $\BDD{U}{\vec{u}} = 1$ iff $\iota(v) = \vec{u}$
for some $v \in U$.  This way, we can represent, \eg, the sets $V_\even$ and $V_\odd$ by
BDDs $\BDD[\even]{V}{\Vars}$ and $\BDD[\odd]{V}{\Vars}$.
The set of edges is given by
a BDD $\BDD{E}{\Vars,\Vars'}$ that ranges over $\Vars$ and $\Vars'$, where
$\Vars'$ is the vector of Boolean variables used to represent a successor
vertex. Finally, the collection of BDDs $\BDD[i]{P}{\Vars}$ represents
the set of vertices with priority $i$.

Set comparison, intersection, union and complement
are easily expressed as BDD operations using equivalence, conjunction, disjunction
and negation.
The pre-image $\pre{\player}{U}$, which is the basis for the APT algorithm, 
can then be encoded using 
existential quantification and the above operations:
\[
\pre{\player}{\BDD{U}{\Vars}}\\
=
(\BDD[\player]{V}{\Vars} \wedge \exists \Vars'. (\BDD{E}{\Vars,\Vars'} \wedge \BDD{U}{\Vars'} ))
\vee
(\BDD[\pnot{\player}]{V}{\Vars} \wedge
\neg \exists \Vars'. (\BDD{E}{\Vars,\Vars'} \wedge \neg \BDD{U}{\Vars'}))
\]
Using the pre-image, the $\player$-attractor 
$\attr{\player}{\BDD{U}{\Vars}}$ and the confined $\player$-attractor
$\attr{\player}{\BDD{T}{\Vars},\BDD{U}{\Vars}}$ 
can then be implemented effectively as a least fixpoint computation.
Note also that the operation $\esc{\player}{G}{A}$, used in lines~5 and~10 of the PP algorithm,
can be implemented using the pre-image operation. The $\Call{DiamondBox}{\,}$
function of the FI algorithm can be implemented in a similar manner:
\[
(\BDD[\even]{V}{\Vars} \wedge \bigvee_i ( \exists \Vars'. (\BDD{E}{\Vars,\Vars'} \wedge \BDD[i]{P}{\Vars'} \wedge \BDD[i]{X}{\Vars'} ) ) )
\vee 
(\BDD[\odd]{V}{\Vars} \wedge \neg \bigvee_i ( \exists \Vars'. (\BDD{E}{\Vars,\Vars'} \wedge \BDD[i]{P}{\Vars'} \wedge \neg\BDD[i]{X}{\Vars'} ) ) )
\]
where, for each $i$, $\BDD[i]{X}{\Vars}$ is the set of vertices represented by approximation
$X_i$ in Algorithm~\ref{alg:FI}.

The operations in the PP algorithm that remain to be encoded involve
the computation of the best escape priority and the computation of
new priority mappings. For the best escape priority we are required to
compute the minimal priority among a set of successor vertices of a
given set of vertices $U$. We split this computation in two parts.
First we identify the set of successor vertices using the operation
$\post{U}$ defined as the union of $uE$, for all $u \in U$; its BDD
encoding is standard: $ \post{\BDD{U}{\Vars}} = (\exists \Vars.
\BDD{E}{\Vars,\Vars'} \wedge \BDD{U}{\Vars})[\Vars' \mapsto \Vars]
$, where $[\Vars' \mapsto \Vars]$ denotes a variable renaming.  
Computing $\bep{\player}{p_g}{A}$, done in line~16 of Alg.~\ref{alg:PP_search}, then can be
implemented by searching for the least value $m$ in the domain of
mapping $p_g$ (assuming that we have BDDs $\BDD[g,i]{P}{\Vars}$ representing the
set of vertices $\{v \in V \mid p_g = i\}$) for which the BDD $\neg\BDD{A}{\Vars} \wedge \BDD[g,m]{P}{\Vars}
\wedge \post{\BDD{A}{\Vars} \wedge \BDD[\player]{V}{\Vars}}$ is
satisfiable. This can be done using a simple iteration.

Finally, the computations of $p_g^*$ in lines~12 and~17 of Alg.~\ref{alg:PP_search} are
simple updates of the function $p_g$ which can be done by iterating over the domain
of the mapping $p$ (resp.\ $p_g$) and computing new BDDs using conjunction and disjunction
of the BDDs given by $p$ and $p_g$. More specifically, for updates of the form
$(p \uplus p_g^{\ge m})[A \mapsto m]$, we first compute the set of vertices
in the domain of $p_g$ that are above $m$:
\[
\BDD[\ge m]{A\textit{\!bove}}{\Vars} = \bigvee \{ \BDD[g,i]{P}{\Vars} \mid i \ge m \}
\]
The updated priority function $p_g^*$ coinciding with $(p \uplus p_g^{\ge m})[A \mapsto m]$, where
$A$ is encoded by BDD $\BDD{A}{\Vars}$ and the set of vertices $\{v \in V \mid p(v) = i\}$ is encoded
by $\BDD[i]{P}{\Vars}$ is then given by:
\[
\BDD[\!\!\!\!\!\!g,i]{P^*}{\Vars} = 
\left \{
\begin{array}{ll}
\neg \BDD{A}{\Vars} \wedge
((\BDD[i]{P}{\Vars} \wedge \neg \BDD[\ge m]{A\textit{\!bove}}{\Vars} )
\vee 
(\BDD[g,i]{P}{\Vars} \wedge \BDD[\ge m]{A\textit{\!bove}}{\Vars}) )
& \text{if $i \neq m$}
\\
\BDD{A}{\Vars} \vee
(\BDD[i]{P}{\Vars} \wedge \neg \BDD[\ge m]{A\textit{\!bove}}{\Vars} )
\vee 
(\BDD[g,i]{P}{\Vars} \wedge \BDD[\ge m]{A\textit{\!bove}}{\Vars})
& \text{else}
\end{array}
\right .
\]

\section{Experimental Evaluation} \label{sec:experiments}

We have implemented all four algorithms in Python,
utilising the BDD package \textsf{dd}\footnote{See
\url{https://github.com/johnyf/dd}, version 0.5.4, by Ioannis
Filippidis} which, next to a native Python BDD implementation,
offers wrappers to C-based BDD implementations such as CUDD, Sylvan
and BuDDy. For the timings we report on for our experiments we rely
on the native Python implementation as our independent experiments
suggest that its performance is comparable to that of CUDD.  Note
that the choice of programming language is secondary given that all
essential time-consuming computations involve BDD creation and
manipulation. We conducted all experiments on a Macbook Pro, 3.5
GHz Intel Core i7 (13-inch, 2017 model, \ie, dual core), with 16 Gb
2133 MHz LPDDR3 main memory, running macOS 10.13.4.

We compare the performance of the implementation of all four
algorithms as described in Section~\ref{sec:implementation}, and,
for reference, we also compare these to explicit-state state-of-the-art
C++ implementations of both the Zielonka and priority promotion
algorithms offered by the \texttt{pbespgsolve} tool of the mCRL2
toolset~\cite{CranenGKSVWW13}.  First, we assess their performance
on random games with different characteristic. Such games, however,
may be poor predictors for the performance of the algorithms on
practical problems. Second, since one of our aims is to assess the
scalability and the speed at which BDD-based solvers can solve
parity games that encode practical verification problems, we also
compare their performance on some of the larger cases of the Keiren benchmark
set~\cite{Keiren}.

\newcommand{\TO}{\ensuremath{\mathbf{\dagger_T}}}
\newcommand{\OOM}{\ensuremath{\mathbf{\dagger_M}}}
\subsection{Random Parity Games}

PGSolver~\cite{FriedmannL09} is a collection of parity game
solving algorithms programmed in OCaml, which includes a tool
\texttt{randomgame} to generate (possibly self-loop-free) random
parity games with a fixed number of vertices.  Running
$\texttt{randomgame~N~P~L~H~x}$ generates a game with \texttt{N}
vertices; each vertex $v$ is assigned a priority that is chosen
uniformly from the interval $[0,\texttt{P}-1]$, and ownership $\even$
with probability $\frac{1}{2}$ and ownership $\odd$ with probability
$\frac{1}{2}$. Each vertex is connected to $d$ distinct neighbours,
where $d$ is chosen uniformly from the interval $[\texttt{L},\texttt{H}]$.
Since the tool generates explicit graphs in the PGSolver format,
we automatically converted these games to BDDs using a binary encoding
for the vertices. 

For our first batch of experiments, we considered both dense and low
out-degree, self-loop-free random games with a high number of distinct 
priorities, varying $\texttt{N} \in \{250,500,1\,000,2\,000,4\,000\}$:
\begin{itemize}
\item low out-degree random games generated using \texttt{randomgame N N 1 2 x};
\item dense random games generated using \texttt{randomgame N N 1 N x};

\end{itemize}
For each class of random games and each value of \texttt{N} we generated 20 parity games.
Table~\ref{tab:random_pgsolver} lists the cumulative runtime (in seconds) for solving these
parity games. We set a timeout of 12\,000 (cumulative) seconds for each experiment;
a timeout is denoted by $\TO$. Experiments requiring over 15Gb memory are aborted; we 
indicate this by $\OOM$. 
\begin{table}[h!]
{\footnotesize
\centering
\setlength{\tabcolsep}{3pt}
\begin{tabular}{l||lllll||llllll}
         & \multicolumn{5}{c}{\emph{low out-degree games}} & \multicolumn{5}{c}{\emph{dense games}} \\
Solver   & 250 & 500 & 1\,000 & 2\,000 & 4\,000      & 250 & 500 & 1\,000 & 2\,000      & 4\,000 \\
\hline
Zielonka & 9   & 53 & 302   & 1\,130 & 3\,463      & 43  & 228 & 1\,004 & $\OOM$ & $\OOM$ \\
PP       & 10  & 63 & 316   & 815  & 3\,048      & 50  & 227 & 1\,039 & $\OOM$ & $\OOM$ \\
FI       & $\TO$   & $\TO$  & $\TO$     & $\TO$    & $\TO$         &  $\TO$  &  $\TO$  &  $\TO$   & $\OOM$ & $\OOM$ \\
APT      & $\TO$   & $\TO$  & $\TO$     & $\TO$    & $\TO$         &  $\TO$  &  $\TO$  &  $\TO$   & $\OOM$ & $\OOM$ \\
\hline
Zielonka (explicit) & 0 & 0 & 0 & 0 & 1 & 0 & 0 & 0 & 0 & 5   \\
PP (explicit)       & 0 & 0 & 0 & 0 & 0 & 0 & 0 & 0 & 0 & 3   \\
\end{tabular}
\caption{Cumulative time (in seconds) to solve the random games for a given value
of \texttt{N}. We set a timeout of 12\,000 seconds (indicated by $\TO$) for the 
cumulative solving time; $\OOM$ indicates out-of-memory.
}
\label{tab:random_pgsolver}
}
\end{table}

We observe that for both types of games, the explicit Zielonka and
PP solvers substantially outperform the symbolic solvers, see
Table~\ref{tab:random_pgsolver}.  Of note is that both APT and FI
fail on all accounts, which we believe is due to the large number
of distinct priorities in the random games, causing both
algorithms to require an excessively large number of iterations.
The symbolic PP implementation outperforms the symbolic Zielonka
implementation for low out-degree games that have large number of vertices and
different priorities, but for smaller number of vertices and small
number of different priorities, Zielonka performs better, but only
marginally so.  For games that consist of more than 2\,000 vertices,
the BDD approach fails dramatically, especially in the case of dense
games.

We can trace the poor performance of the symbolic solvers to the
overly complex and unstructured BDDs underlying the random graphs.
These unwieldy BDD representations are caused by the lack of structure
provided by the binary encoding of vertices and edges, leading to
a poor compression ratio: typically, BDDs remain concise if two
adjacent vertices (\ie, vertices related by $E$) only differ in their
representation by a relatively small number of bits. The binary
encoding, however, does not guarantee this. Variable-reordering heuristics
may help in some cases but are not guaranteed to do so. As a result, the BDDs
themselves require an excessive amount of memory (as witnessed by
the out-of-memory issues in modest-sized dense random graphs), and
operations on these BDDs lead to excessive running times. As such,
these experiments may not be good predictors for running times of
the symbolic algorithms on games encoding practical verification
problems.  However, they do suggest that FI and APT both suffer
significantly from a poor compression ratio. Zielonka and PP, on
the other hand, are (relatively) less susceptible to a poor compression
ratio.\medskip

For our next batch of random experiments, we repeat the experiments
conducted in~\cite{StasioMPV16}. In that work, the authors observe
that the APT algorithm outperforms Zielonka's algorithm for games
in which the number of vertices is exponential in the number of
priorities in a game, as long as the base is sufficiently large:
they show that APT outperforms Zielonka for $\texttt{N} \approx c^\texttt{P}$, for
$c = e$ or $c = 10$, but for $c = 2$, Zielonka outperforms APT.
Games in which the number of vertices is exponential in the number
of distinct priorities are typical of games encoding practical verification
and synthesis problems. 
We generate dense random games with a priority/vertex ratio given
by the relation $\texttt{N} \approx c^\texttt{P}$, for $c \in \{2,e,10,13\}$, using
$\texttt{randomgame~N~P~1~N~x}$; \ie, we extend the experiments of~\cite{StasioMPV16}
by also considering $c = 13$. We again generate 20 random games per
experiment, and set a timeout at 12\,000 seconds (cumulative),
see Table~\ref{tab:random_pgsolver2}.

\begin{table}[h!]
{\footnotesize
\centering
\setlength{\tabcolsep}{3pt}
\begin{tabular}{l||lllll||lllll|lll|lll}
         & \multicolumn{5}{c}{$\texttt{N} = 2^\texttt{P}$ \emph{games}} & \multicolumn{5}{c}{$\texttt{N} \approx e^\texttt{P}$ \emph{games}} & \multicolumn{3}{c}\emph{$\texttt{N} = 10^\texttt{P}$ \emph{games}} & \multicolumn{3}{c}\emph{$\texttt{N} = 13^\texttt{P}$ \emph{games}}\\
Solver   & 128 & 256 & 512 & 1\,024 & 2\,048      & 21 & 55 & 149 & 404 & 1\,097    & 10 & 100 & 1\,000 & 13 & 169 & 2197\\
\hline
Zielonka & 9   & 33   & 127   & 559  &  2\,203  & 1 & 2 & 14 & 88     & 646     & 0 & 7 & 514   & 0 & 18 & 3099\\
PP       & 10  & 37   & 132   & 543  &  2\,355  & 1 & 2 & 15 & 91     & 717     & 0 & 8 & 544   & 0 & 19 & 4299\\
FI       & 348 & 2\,212 & $\TO$ &$\TO$ & $\TO$  & 1 & 6 & 89 & 955    & 10\,695 & 0 & 6 & 530   & 0 & 12 & 2912\\
APT      & 173 & 1\,092 & 8\,195  &$\TO$ & $\TO$  & 1 & 4 & 61 & 659    & 7\,526  & 0 & 5 & 545 & 0 & 13 & 3982\\
\hline
Zielonka (explicit) & 0 & 0 & 0 & 0 & 1 & 0 & 0 & 0 & 1 & 4 & 0 & 0 &3 & 0 & 0 & 17 \\
PP (explicit)       & 0 & 0 & 0 & 0 & 1 & 0 & 0 & 0 & 1 & 4 & 0 & 0 &3 & 0 & 0 & 17 \\
\end{tabular}
\caption{Cumulative time (in seconds) to solve the random games for a given value
of \texttt{N}.  A timeout (a cumulative time of 12\,000 seconds or more) is indicated
by $\TO$. 
}
\label{tab:random_pgsolver2}
}
\end{table}

Table~\ref{tab:random_pgsolver2} illustrates that Zielonka has a slight edge over PP.
Comparing the results in Table~\ref{tab:random_pgsolver2} to the
experimental outcomes in~\cite{StasioMPV16}, we find some interesting
differences. First, whereas in~\cite{StasioMPV16}, the APT algorithm
is found to significantly outperform Zielonka's algorithm for $c =
e$ and $c = 10$, our results do not provide such clear indications.
We do see that for $c = 10$ and $c = 13$, the running times of the
APT algorithm becomes more in line with that of Zielonka and PP.
The behaviour of FI illustrates a more pronounced correlation between
the number of vertices and the number of priorities in a game,
suggesting that in a symbolic setting, the FI algorithm may actually
perform quite well on games that encode verification problems that
typically result in only a few distinct priorities. We test this
hypothesis by comparing the performance on a number of (large)
parity games originating from model checking problems, mostly taken from
the Keiren Benchmark set.

\subsection{The Keiren Parity Game Solving Benchmark Set}

In~\cite{Keiren}, Keiren describes and provides a set of parity
games that originate from over 300 model checking problems, originating
from 21~specifications. The main obstacle in reusing
the data set is that the games are encoded as explicit graphs in
the PGSolver format. This means that most of the structure that a
BDD solver can typically exploit for compactly representing the
game graph is lost. As we concluded from the random games generated
by PGSolver, a binary encoding of the sets involved leads to severe
performance degradation of the solvers, and given the size of the
graphs, there is little hope the running times of our algorithms
on such encoded BDDs provide meaningful information. We have coped
with this by generating BDDs from the original specifications for
several of the model checking games included in the Keiren benchmark
set.\footnote{To this end, we
developed a special-purpose \texttt{pbes2bdd} tool for converting (a fragment of) parameterised Boolean
equation systems in Standard Recursive Form, containing only Boolean data parameters, to BDDs.
The tool currently does not guarantee totality of the generated parity game; instead,
it relies on the user to guarantee that this is the case.}
Note that the conversion of the original specifications to
BDDs is not straightforward: most games stem from model checking
problems for system models that employ (unbounded) lists, natural
numbers, \emph{etcetera}, for which no trivial bounds can be
established. The results can be found in Table~\ref{tab:keiren};
here, $\OOM$ indicates that the computation was aborted because it
needed more than 15Gb of memory.

\begin{table}[h!]
{\footnotesize
\centering
\begin{tabular}{l|p{5.7cm}p{0.6cm}p{1.5cm}p{1.2cm}p{1.0cm}p{1.0cm}p{0.3cm}}
Model & Property & $\#$prio & Zielonka (sym/exp) & PP (sym/exp) & FI & APT  & $\#\Vars$\\
\hline
Chatbox       & Absence of Deadlock                      &2 & 0.04/0.02 & 0.05/0.01  & \textbf{0.00} & 0.01  &  18\\
              & It is possible to JOIN infinitely often  &2 & 0.02/0.02 & 0.02/0.01  & \textbf{0.00} & 0.01 &   18\\
              & Invariantly LEAVE can follow JOIN        &3 & 0.04/0.07 & 0.05/0.02  & \textbf{0.01} & 0.02 &   18\\
Onebit$_2$    & Infinitely often send and receive        &3 & 0.59/0.14 & 0.58/\textbf{0.04}  & 0.64 & 0.69     &   26\\
              & Absence of Deadlock                      &2 & 0.16/0.11 & 0.19/\textbf{0.03}  & 0.04 & 0.09 &   26\\
Onebit$_8$    & Infinitely often send and receive        &3 & 3.30/24.17 & 3.28/4.49  & 3.36 & \textbf{3.19} &   38 \\
              & Absence of Deadlock                      &2 & 0.86/22.70 & 1.02/6.82  & \textbf{0.19} & 0.47   &   38 \\
Onebit$_{32}$ & Infinitely often send and receive        &3 & 10.93/$\OOM$ & 10.68/$\OOM$ & 10.22 & \textbf{9.81} &   50 \\
              & Absence of Deadlock                      &2 & 2.42/$\OOM$ & 3.16/$\OOM$ & \textbf{0.52}  & 1.65  &   50 \\
Hesselink$_4$ & Absence of Deadlock                      &2 & 1.97/$\OOM$ & 1.87/$\OOM$ & \textbf{1.21}  & 1.31   & 60 \\
              & Cache consistency                        &2 & 3.47/$\OOM$ & 3.55/$\OOM$ & \textbf{1.55}  & 1.93   & 61 \\
Dining$_{13}$ & Absence of Deadlock                      &2 & 1.45/8.40 & 1.46/7.68 & 0.20 & \textbf{0.19} & 89 \\
              & Aristotle can eat infinitely often       &2 & 7.58/3.44 & 8.00/\textbf{1.18} & 4.24 & 4.03 & 89 \\
              & Aristotle can, and thus will eat infinitely often 
                &3 & 1.69/10.35 & 1.56/4.74 & 0.43 & \textbf{0.37} & 90 \\
\multicolumn{8}{c}{\phantom{empty}} \\
 \multicolumn{2}{r}{Cumulative:} &  & 34.52/$\OOM$ & 35.47/$\OOM$ & \textbf{22.61} & 23.77 & \\
\end{tabular}
\caption{Running times (in seconds) for selected model checking
problems; $\OOM$ indicates an out-of-memory error; $\#\Vars$ indicates
the number of Boolean variables needed to encode the set of vertices.}
\label{tab:keiren}
}
\end{table}
We again compared our BDD-based implementations of the four parity
game solving algorithms to the explicit Zielonka and PP solvers of
the \texttt{pbespgsolve} tool\footnote{Of note is the fact that for our
explicit solvers, we only compute the solution to the vertices
reachable from a designated vertex (typically the vertex that
indicates whether the modal $\mu$-calculus formula holds for the
initial state of a system); we have excluded the (typically significant)
time required for constructing this reachable subgame and only list the time
required to solve the resulting (sub)game. The game that is
solved by the explicit solvers can thus be orders of magnitude smaller
than the games solved by our symbolic solvers, as these do not 
conduct a reachability analysis.} (indicated by the `exp' fields in Table~\ref{tab:keiren}).

First, observe that all four symbolic solvers perform quite well
on all games, sometimes significantly outperforming the two explicit
solvers.  In particular, none of the symbolic solvers run out-of-memory;
in fact, memory usage for all games stays
well below 500Mb. This sharply contrasts the explicit solvers,
which, for the larger games that can be solved consume 8Gb or more. We furthermore
note that the runtime performance of the symbolic solvers appears
to be more robust than that of the explicit solvers.

Second, we observe that both FI and APT clearly outperform Zielonka
and PP.  The PP and Zielonka algorithms perform comparably on these
games, with Zielonka having an insignificant edge over PP; in this
regard, the outcomes of these experiments are largely similar to
the collective outcome of our experiments using random games.

\section{Discussion and Conclusions} \label{sec:conclusions}

We studied four algorithms for solving parity games, \viz Zielonka's
recursive algorithm, the recently introduced Priority Promotion (PP)
algorithm, the Fixpoint-Iteration (FI) algorithm and the APT algorithm,
and compared the performance of their symbolic implementations.

Our results indicate that, overall, the symbolic implementations
of Zielonka's algorithm and the PP algorithm have very similar
performance characteristics. The experiments using random games
indicate that Zielonka's algorithm performs slightly better than
PP for games with a small number of priorities, and for games in
which the number of vertices is exponentially larger than the number
of distinct priorities, but the differences are not significant in
most cases.  The same observation applies to the cases taken from
Keiren's benchmark set.

The symbolic implementations of the APT and FI algorithms both
perform poorly on most of our random games. There are indications
that also in the symbolic case, these algorithms are more
suited for games in which the number of vertices is exponentially
larger than the number of distinct priorities, but the results
are not as pronounced as reported in~\cite{StasioMPV16} for the
explicit case. We believe that part of this difference is due
to the susceptibility both algorithms demonstrate on the size of the
BDDs.  This is further supported by the cases taken from Keiren's
benchmark set, in which the BDDs remain rather small and both
FI and APT outperform Zielonka and PP, but further research is
needed to confirm this hypothesis and to see if there are ways
to sidestep such issues.

While our experimental results on the Keiren benchmark set show
that both FI and APT beat Zielonka and PP on games stemming from
verification problems, their unpredictability in the presence of
large BDDs makes them less reliable for solving such problems as one cannot
upfront predict the size of BDDs in practice.
Given that all solvers require at most seconds for even the largest
models, containing up to $2^{90}$ vertices, any of the four algorithms
would work in practice, and one may therefore favour the more robust
PP or Zielonka algorithms in practice.

Finally, comparing the performance of our symbolic Zielonka and PP solvers to the
explicit solvers, we find that there are only few cases in which
the explicit solvers have an edge over the symbolic solvers. More
importantly, the explicit solvers are unable to solve games which
are readily solved by the symbolic solvers, indicating the practical
significance of BDD-based solvers.

\bibliographystyle{eptcs}

\end{document}